# Event-based Data Format Standard (EVT+)

Ver: 0.7d

08/18/2025

Contributors:

- Dr. Jonah P. Sengupta, DEVCOM ARL, jonah.p.sengupta.civ@army.mil
- Dr. Mohammad Imran Vakil, AFRL-RYDH, mohammad.vakil.1@us.af.mil
- Mr. Thanh M. Dang, DEVCOM ARL, thanh.m.dang.ctr@army.mil
- Mr. Ian Pardee, NSWC Crane, ian.a.pardee.ctr@us.navy.mil
- Mr. Paul Coen, NSWC Crane, paul.d.coen.civ@us.navy.mil
- Dr. Olivia M. Aul, NSWC Crane, olivia.m.aul.civ@us.navy.mil

# Acronyms

AEDAT4 – File format created by iniVation AG for event sensor data storage.

AER – Address Event Representation

ASCII – American Standard Code for Information Interchange

DAC – Digital to Analog Converter

EBS – Event Based Sensor / neuromorphic sensor / dynamic vision sensor / event-based camera / event camera

EVT – EVent Type, data standard name created by Prophesee

FoV – Field of View

FPA – Focal Plane Array

FPGA – Field Programmable Gate Array

IMU – Inertial Measurement Unit

LLSB – Lower-lower significant bits

LSB – Least Significant Bit(s) – Refers to the lowest power bit(s) in a binary representation.

MSB – Most Significant Bit(s) – Refers to the highest power bit(s) in a binary representation.

ROI – Region Of Interest

ROIC – Read-Out Integrated Circuit

SOTA – State Of The Art

TS – Time Stamp

TS LSB – Time Stamped Least Significant Bit

TS MSB – Time Stamped Most Significant Bit

# Contents

# LIST OF TABLES

# 1 Introduction

Event-based sensing (EBS) hardware is quickly proliferating while finding foothold in many commercial, industrial, and defense applications. At present, there are a handful technologically mature systems, most notably from Prophesee and iniVation, which produce data streams with diverse output formats. In the near future it is anticipated there will be American vendors who offer new sensor hardware which could also yield unique data schema that are not aligned to past efforts. Thus, due to the relative nascent nature of the technology and its potential for widespread use in a variety of applications, it is an opportune time to define a standard for this class of sensors' output data. The intent of this document is to identify and provide a standard for the collected EBS streaming data. The main objective of the standard is to be sensor agnostic, incorporate some of the current sensor configurations and modalities, and account for the developing configurations and modalities. The intent is also to leave enough place holders and space in the standard for future variations that may develop as EBS technology matures.

This is a living document and as such will have more than one version. This document will be updated as needed or required due to tech advances in the field. Since EBS are already in use and producing data, it is prudent to have a standard that can be applied to both the existing hardware and prospective sensors that will be produced/collected moving forward by both United States federal and commercial entities. Field data collection is always a significant issue in modeling and simulation as well as algorithmic development for scene analysis, since no dataset can account for every slight deviation in the sensor field of view. Collecting samples can be especially costly for operational/field scenarios, limiting the environmental conditions and number of varying targets. To mitigate this dearth of data, a streaming data standard will enable efficient data swap between entities for creating models and algorithms for EBS while minimizing the need for costly and time-consuming redundant collection efforts.

The data breakdown will be kept in a standard format of Header/Metadata and streamed data. The header data will vary depending on the hardware configuration of the sensor. Some prospective EBS hardware may provide the capability to collect intensity images in tandem with the event-based data; this will also be identified among other parameters in the header which is described in the following sections. The header is 256-bits and provides space for these various vendor parameters, in addition to providing critical space to append sensor endpoint information such as analog bias values and on-chip metrology (temperature, bandwidth, illumination).

The data produced by an EBS imager is complex as it provides multiple parameters and can be in different configurations. To mitigate issues with data, an 8-bit datum header is introduced which allows more clarity and precision for varying data types dependent on modality and hardware configuration. It also provides ample room for users to append more data types depending on hardware endpoints or desired applications. The data section of this document will provide more details on these various payload structures. It will also be prepended with typical operational situations which may leverage the full swathe of support packet types.

The structure of the data payload is greatly influenced by the EVT3 standard [1] outlined by Prophesee, commercial vendor of EBS software and hardware. The EVT3.0 standard is a 16-bit protocol which supports typical serialized address events, 32-bit timestamps, and vectorized representations. This latter operation allows for optimized output bandwidth when there is a large event throughput. Fitting space for both the full serialized (Y then X) protocol and the vectorized compression allows the standard to optimize bandwidth with respect to the output event throughput. While innovative and fitting for current commercial IP, EVT3.0 only has 16 payload types, no support for other data payloads, finite timestamp resolution, and no header space for critical metadata that is invaluable for software developers and sensor integrators. As such, this standard, coined EVT+, is a reimagining of EVT3.0 and provides a flight path to support future sensors which may be multi-modal (meaning capable to stream images and events simultaneously).

## 2 Standard Overview

Like most other protocols [2], the full standard can be broken down into two distinct parts: the data header and the data payload. The former provides crucial metadata about the ROIC, the FPA, the firmware, and the configuration of the system on a per-data-frame basis. This is critical as some user applications may have the ability to reconfigure the device during runtime. The header decoding format is further divided into minimum required bits, a flexible number of bits dependent on user defined variables, and a list of pointers for specific time intervals which in turn will depend upon number of event activations per frame.

The latter element, the data payload, consists of a large vector of 1 of the 14 data payload variants supported by the standard. At present, the standard has space for 255 data types so there is a large sandbox for developers to insert new payloads into the format. Each data frame only supports one of the 7 data modalities for the image data payload. A brief overview of the current supported variants is given in the Data Modality subsection in the next section. This is different than the 14 data types as these data modalities express how the visual data is communicated. These modalities can comprise of multiple data types. For instance, pure event readouts require the TS MSB, EVENT Y, and EVENT X data types for full communication. More information on the header and data payload types are thoroughly described in the following sections.

# 3 Header Descriptions

## 3.1 Header Decoding Table

| # | Header Variable | Description | Bits Assigned (Notional) |
|---|---|---|---|
| 1 | Header ID | Identifies that the following data is correlated to the header. | 8 |
| 2 | Epoch Timestamp | Epoch Timestamp slot to be filled in by host system | 64 |
| 3 | Global Time Stamp | Synchronization of data | 24 |
| 4 | Sensor Modality | Single, dual, quad mode | 3 |
| 5 | Data Modality | If not single mode, then data modality | 3 |
| 6 | Number of Frames | Number of frames streamed | 18 |
| 7 | Rows | Dimension of the FoV | 16 |
| 8 | Cols | | 16 |
| 9 | Reserved | Reserved bits For Future Use Initialize to zeros | 64 |
| 10 | Sensor/ROIC Manufacturer/Model | 32 ASCII characters to enter the information | 256 |
| | | **Required Header bits** | **472 bits** |
| 11 | **User Defined Hdr Size** | **# Of 32-bit words in User Defined Var** | **16** |
| | User Defined Word 0 | Available bits for user/vendor to fit ROIC parameters w/info about DAC, FPGA etc. | 32 |
| | User Defined Word 1 | | 32 |
| | … | | 32 |
| | User Defined Word N | | 32 |

| | Pointers to Timestamps Size | # Of Pointers | 32 |
|---|---|---|---|
| | Time Ptr Increment | # Of Microseconds | 32 |
| | Ptr0 | Pointer to start time | 32 |
| | Ptr1 | Pointer to start time plus 1 increment | 32 |
| | Ptr2 | Pointer to start time plus 2 increments | 32 |
| | … | … | 32 |
| | PtrN | Pointer to start time plus N increment | 32 |

*Table 1 Proposed Header for EBS data*

## 3.2 Header Member Description

### 3.2.1 Header ID (8 bits)

The ID tag for the header section of the data – this is inset by the encoder such that the decoder can rapidly identify the header section and then subsequently read 32 bytes of the data. All data members of the header are byte aligned (outside of the sensor/data modality sub-section) thus allowing for rapid masking. Other than the header, all other data payload types are 4 bytes (32-bits).

### 3.2.2 Epoch Timestamp (64 bits)

This is reserved for an epoch time stamp to be filled in by the host system. Initially, this is initialized to all 0's, to allow for flexibility between hosts. This will act as a stamp from the host machine to acknowledge "data has been received starting at this time according to the system."

### 3.2.3 Global Timestamp (24 bits)

A time stamp synchronizing the EBS system with one or more external sensors (e.g., IMUs) for correlation of data collected by both sensors. The global time stamp will mark the start of data collection by the EBS system. This is different from the internal time stamp used for each event or frame.

### 3.2.4 Sensor Modality (3 bits)

An EBS system may provide, in addition to event data, the capability to stream intensity data like a traditional framing imager. This capability makes it a dual mode sensor. Future variation may include polarization and other features. There are 3 bits assigned for sensor modality providing 8 different modes for future additions. An example encoding scheme may be:

0. Event imager only (Prophesee/iniVation dvXplorer IP)
1. Event + Framing sensor (iniVation DAVIS IP)
2. 2-color event + framing sensor

### 3.2.5 Data Modality (3 bits)

A single mode EBS system may provide event data as a function of change in intensity (p), time of change (t), and location (x, y) in the field of view.

$$\text{Data} = d(x, y, p, t) \quad (\text{eq. 1})$$

A full framing imager streams the n bit intensity, pixel by pixel in its active field of view. The data modality parameter indicates the expected output data type. A dual mode system may be configured to output either event data or active pixel intensity data. As in the case of sensor modality, there are 3 bits assigned for data modality providing 8 different modes for future additions. The seven data modalities are:

1. Frames of intensity data
2. Event integration images: 2D frame where each pixel is number of events detected over a fixed amount of time or event activity. These parameters are set by the user
3. Event dt images: the instantaneous time between the most recent events in the pixel.
4. Event single payload: pure Address-Event (AE) representation (see eq. 1 above)
5. Mixed single payload: AE representation with each event accompanied by the intensity $i$ of that pixel ($ae\_m = i_m(x,y)$) where $i_m$ is the 32-bit intensity of that pixel).
6. Vectorized event payload: address events are encoded via 1-hot vectors per polarity where the x (column) address is resolved using a X-base value and the position of asserted ones in the vector.
7. Vectorized mixed payload: asserted 1-hot members of the vectors are followed by intensity values which correspond temporally to the events.

### 3.2.6 Number of Frames/Datum (18 bits)

There are 18 bits assigned to this parameter. 2^18 allows up to 262,144 frames or datum to be streamed in each mode. Datum can be either a frame or type of event. Types of events include pure events, mixed events, vectorized events, or mixed vectorized events.

### 3.2.7 Rows, Cols: frame Size in Field of View (FoV) (16 bits x 16 bits)

Frame size is the product of active rows and columns. The system parameters may allow a subset of row and columns to provide a small FoV or use the entire detector array to have a full FoV. 16 bits each are assigned for rows and columns providing an array size of 65,536 by 65,536 accounting for current and future array sizes.

### 3.2.8 Sensor/ROIC Manufacturer/Model (256 bits)

This field provides the user/operator with 32 ASCII characters [3] to define the system they employed for data collection. It also eliminates the need for a table for specific bit assignments of various Sensor model/ROIC numbers/ manufacturers. The user can input relevant info in the 32 ASCII fields for their particular system.

### 3.2.9 User Defined Header for User Defined Variables (16/32 bits)

User defined header utilizes 16 bits to specify how many 32 bits words will follow. Each ROIC/sensor may provide the ability to program how the image is collected e.g. the active FoV in the detector array, threshold for change detection etc. User Defined Variable parameter displays the values used for settings each time an image is acquired. This 32-bit parameter can be divided into many parts to denote flags, parameters, enables, resets, internal states, etc. The total parameter space is flexible and is defined in the User Defined Header. These parameters may encompass:

- Digital-to-analog converter values used to bias the event-based pixel circuit or set the requisite supply levels
- Region-of-interest(s)
- Vectorization bin-times or number of events per-bin
- Event dt type: opposite polarities, same polarities
- Desired output mode for payload (see header value #2 for information on different payload types).
- Maximum throughput, digital signal processing values (temporal filtering…)
- Timestamp resolutions

# 4 Pointer to Timestamps (32 bits)

This parameter provides resolution to the saved data in terms of Timestamp packets. It provides the number of Timestamp Pointers, 32 bits each, which are in turn used to define the time difference between current and next pointer. This provides a user with the capability to jump to the specific time interval to retrieve event data without going through the data stream sequentially.

# 5 Data Payload Breakdown

The data payload follows the propagation of the header data to the processing endpoint. The structure of these different data types closely follow the format outlined by the 16-bit EVT3 standard. EVT+ extends the vectorization support of EVT3 to also allow for the transmission of non-event data. EVT+ also allows for extended timestamp resolutions. Here are some of the key guiding principles to the data payload supported in EVT+:

## 5.1 Sharing the timestamps and row data:

**Go per-word modular to amortize the timestamp bits and reduce bit costs (many events and intensities may share the same time stamp), same concept applies PER row as well (many events WILL share the same row).**

In the first case, a timestamp can be shared if all the events transmitted were received in the same temporal period. Under high rates and fine periods, this assumption may collapse, but in practice, the finite latency of the pixel and interfaces have shown that microsecond resolution is more than sufficient. Thus, if the timestamps are the same between two events, they can be shared and data can be saved. In the second scenario, it is also true that event activations often occur in close spatial proximity, thus sharing the same row or column. In that case, assuming the events are sampled in the same window in time, the bits allocated for encoding the row address can be shared. Moreover, many modern AER protocols adopt a row-based readout scheme where rows are selected by an arbiter and all asserted pixels along the columns are read out. Therefore, the readout protocol supports this row-based compression scheme as well, especially if simultaneous activation or high-load is present in the array.

## 5.2 Clipping intensity information

**Only send the intensity image data from pixels that have fired, thus implicitly compressing frame information**

Similar to the AER data format developed by iniVation, EVT+ will also propagate the digital value from the traditional image sensor that is either housed in the same camera body or originates from a hybrid pixel on the ROIC. However, the "MIXED" data types in the standard will only send this digital value from pixels that have sent an event and share an address with the imaging pixel. Thus, the intensity space is sampled on an event-based basis and allows for efficient scene compression at a protocol level. Moreover, this provides the user the ability to fully reconstruct temporal contrast targets without needing to read the whole frame through the camera interface. In practice, this corresponds to the intensity values on the edges of moving targets OR the intensity value of a temporally evolving static target. In the former, the inner values of this target can be populated after querying the camera using an ROI that is informed by the edge information.

## 5.3 Extending the data word:

**Going 32 bits (from 16 bits in EVT+) allows for 8-bit datum header (more word space), 32-bit total intensity value, potential for 40-bit timestamp, 32-bits of vectorization.**

| | Word Description | Datum Header (8-bits) | | | | | | | | Data (24-bits) | | |
|---|---|---|---|---|---|---|---|---|---|---|---|---|
| | | | | | | | | | | | | |
| Non-vectorized Event | Timestamp_high | 0 | 0 | 0 | 0 | 0 | 0 | 0 | 1 | TS MSB (24-bits) | | |
| | Y_address | 0 | 0 | 0 | 0 | 0 | 0 | 1 | 0 | Y-address (16-bits) | | TS LSB (8-bits) |
| | X_address (ON) | 0 | 0 | 0 | 0 | 0 | 1 | 1 | 0 | X-address (16-bits) | | TS LLSB (8-bits) |
| | | | | | | | | | | | | |
| Vectorized Event LLSB = 0 | Timestamp_high | 0 | 0 | 0 | 0 | 0 | 0 | 0 | 1 | TS MSB (24-bits) | | |
| | Y_address | 0 | 0 | 0 | 0 | 0 | 0 | 1 | 0 | Y-address (16-bits) | | TS LSB (8-bits) |
| | VEC_X_ON_MSB | 0 | 0 | 0 | 0 | 1 | 0 | 0 | 0 | Starting X-address (16-bits) | | First 8 pixels 1-hot code (8-bits) |
| | VEC_X_LSB | 0 | 0 | 0 | 0 | 1 | 0 | 1 | 0 | Next 24 pixels 1-hot code (24-bits) | | |
| | | | | | | | | | | | | |

Even though the number of bits per data is doubling, the increased resolution in intensity dynamic range, large space for future data types, and state-of-the-art temporal resolution are immediate benefits. To the first benefit – many modern digital ROICs support well depths exceeding 16-bits with some extending 32-bits. Thus, it is reasonable to expect that support for larger well depths in event-based operation will be needed. Secondly, there are only 16 data types supported now (11 described in the Prophesee documentation, with 5 reserved), but camera vendors of the future may want to integrate other data payloads, such as inertial sensors or magnetometers, etc. With 8 bits, there are 240 additional datums that can be described. Thirdly, extending to 32 bits allows for the TS MSB data to hold 24 bits, while the EVENT Y (row address) data to hold 8 bits. Already this matches the 32 bits supported in EVT3. However, if sending serialized event data, the EVENT X datum has 8 more bits that can be used for the LLSB (lower-lower significant bits) of the timestamp. Thus, streaming data in this manner can get the user 40-bits of temporal dynamic range. With a temporal resolution of 1 microsecond, this dynamic range allows for over 12 days of record space. Conversely, 40-bit range with an 18 minute record space equates to a 1ns resolution: something 3 orders of magnitude finer than SOTA. Finally, the potential bandwidth growth with an increase in datum bits from 16 to 32 bits is curtailed by improved vectorization support. This will be detailed more in the following subsection, but 32-bits allows for two datum to one-hot encode 32 columns of pixels. This increases the probability that a vector will include more than enough events to buy down the cost of serializing data. It also is a 2.75x increase on the vector sampling space seen within the EVT3 protocol (32 vs. 12 columns).

## 5.4 Vectorization – Optional Feature:

**If the vendor has implemented a way to vectorize on their readout or in the FPGA, bandwidth can be greatly reduced under high load cases.**

Vectorization is encoding pixel addresses in a one-hot format in combination with a "root" address that is passed prior. For instance, locations 32, 34, 36, 38 can be propagated with 0010 0000 followed by 0101 0101, instead of 0010 0000, 0010 0010, 0010 0100, 0010 0110. In this scenario, we can compress from 8 bits/event to 2 bits/event. A clear crossover point occurs when:

$$N_{bits,root} + N_{bits,offst} * [abs(X_{root} - \max(\boldsymbol{X_{vec}})) \% N_{bits,offst} + 1] = N_{bits,root} * |\boldsymbol{X_{vec}}|$$

where $N_{bits,root}$ is the number of bits used to encode the event address, $N_{bits,offst}$ is the number of bits used to one-hot encode the vectorized events, $X_{root}$ is the address of the root address for the vector, and $X_{vec}$ is the set of addresses of activated pixels within the vector. The left side of the equation is the number of bits for encoding this vector of event addresses in a vectorized format while the right is the naïve method of sending binary values for each address.

It is quickly evident that vectorizing events in close proximity to the root address is much more efficient than sending each address. However, as the maximum address in this vector grows while the size of the set maintains, it becomes more efficient to send individual addresses. In EVT+, $N_{bits,root}$ is 16-bits while $N_{bits,offst}$ is essentially 0 (8-bits are allotted in the base datum where that is used by 8-bits of timestamp in the pure EVENT X, thus you are trading off time for spatial compression) if the max event in the vector is within 8 column addresses of the root. However, if that exceeds 8, then $N_{bits,offst}$ = 24. With only one other event in the outer group of the vector spanning 32 columns, it is slightly more efficient to send individual addresses and more temporal resolution is afforded (40 bits vs. 32 bits). However, three events in the vector within 32 pixels is more efficiently transmitted via the vectorized format (40 vs. 48 bits). More thorough graphical analysis of this tradeoff with different event activation profiles, bit depth, and offsets is critical, but outside of the scope of this standards document.

Vectorization can be achieved by integrating events over small increments of time or activity. For instance, the readout can track the number of events per pixel over a small temporal period. Anything greater than 0 will be encoded as a 1 and then ultimately transmitted to a control block. This would use the above relation to decide if the local activity in the one-hot vector merits vectorization or if serialized transmission is preferable. Alternatively, this integration method could adapt the temporal period and vectorize based on event activity (integrate until 100 events have been read).

In the end, the user would sacrifice the fine-grained temporal precision offered by serialized events and the potential to achieve a 40-bit timestamp with the pure EVENT X data type. For instance, if the temporal period needed for high-performance vectorization is 10ms, but the temporal precision can go down to 10ns, then there is a 1e6 loss in temporal resolution. However, for high event rates, the integration period could be small enough that the loss in temporal resolution is not noticeable from an application level.

## 5.5 Typical Operation Modes:

The following lists describe the four operation modes in the standard that support the transmission of events (the other three propagate frames). All lists are ordered with respect to the temporal sequence the data arrives at the decoder. Each shows how data will be typically propagated in each scenario.

### 5.5.1 Baseline event mode

This mode only sends event data, something that can be directly compatible with the Prophesee sensors or the iniVation dvXplorer IP. In summary, after sending the payload header, the timestamp MSB (24 bits) is sent, followed by the row address (EVENT Y) from which there are events, and ending with the column addresses of pixels that have asserted in that row. Each row and column datum has additional timestamp information such that when each address event is ultimately decoded, the final timestamp data will be 40-bits.

1. HEADER (stipulating a pure event mode transmission)
2. TS MSB
3. EVENT Y – 0 (TS LSB presides within)

4. EVENT X - 0 ON/OFF (TS LLSB resides within)
5. …………
6. EVENT X – M ON/OFF (M$^{th}$ event in row 0)
7. ….
8. EVENT Y – N (this will always be sent as long as there are events that belong to this row)
9. ….
10. EVENT X – M ON/OFF (M$^{th}$ event in row N)
11. ….
12. TS MSB (Tick = this will only be sent if the TS MSB has changed from line item 2).
13. ….

### 5.5.2 Mixed Event mode (configured a priori or by user)

In this modality, the camera is sending serialized events which also house the intensity value from a corresponding pixel in an imaging array. This intensity is encoded into a 32-bit digital value which is spread across the lower 8 bits of the MSB datum and the lower 24-bits of the LSB datum. Once these values are decoded by the processing endpoint or frame grabber, the user will have 32-bits of intensity data per address event. Thus, the user can efficiently reconstruct the image given the event activity.

1. TS MSB
2. EVENT Y - 0
3. MIXED X ON/OFF MSB - 0 (event X address from event 0 AND the 8 MSB from the intensity data)
4. MIXED X LSB – 0 (lower 24-bits of the intensity data belonging to the same address of the event)
5. ………
6. MIXED X ON/OFF MSB - M
7. MIXED X LSB - M
8. …..
9. EVENT Y – N
10. …..
11. TS MSB (tick)
12. ….

### 5.5.3 Pure Vectorized event mode (can be *a-priori* or adaptive based on ROIC or FPGA)

This modality assumes that the FPGA or ROIC is vectorizing the event data with respect to some integration or readout scheme. Data can be ultimately decoded using the column address, datum type which indicates event polarity, and the one-hot encoded data vectors. These LSB vectors can be daisy chained if the vectorization controller is advanced enough to recognize if the salient activity within a row merits continuing the data OR if a new VEC X MSB needs to be sent. The following example shows both cases.

1. TS MSB
2. EVENT Y – 0
3. VEC X ON MSB – 0 (this datum houses the vector root address and the first 8 column address)

4. VEC X LSB – 0,0 (the LSB datum follows directly after, this allows the decoder to also correlate polarities. To note, OFF event propagation would require another MSB datum to be sent EVEN if the ON/OFF vectors overlap in space).
5. VEC X LSB – 0,1 (if the vectorization controller can support, another 24 columns can be sent, which one-hot encoded event activation. The column addresses and polarities are decoded in this daisy chain case by recursing back to the root address and tracking how many LSB datum have been sent).
6. ……
7. VEC X LSB – 0,M ($M^{th}$ daisy chained LSB datum)
8. VEC X OFF MSB - 1 (this VEC X OFF root address could be the same as the one encoded in 3, but it can also be different. This datum also encodes the activation of the adjacent 8 pixels. In this case, the next vector of OFF events is far enough removed from this root address such that it is split into another VEC X OFF MSB word: VEC X OFF MSB 2).
9. VEC X OFF MSB - 2…

### 5.5.4 Mixed Vectorized mode

The final supported event-based transmission mode is the mixed vectorized mode. In this state, the camera sends the events in vectorized representation. However, the intensity from each pixel asserted in the one-hot vector is also sent following the transmission of the vectorized event data (VEC X ON/OFF MSB, VEC X LSB…).

1. TS MSB
2. EVENT Y – 0
3. VEC X ON MSB - 0
4. VEC X LSB – 0,0
5. VEC X LSB – 0,1 (56 pixels total)
6. VEC X INTENSITY MSB – 0 + 0 (first asserted pixel in vectorized word)
7. VEC X INTENSITY LSB – 0 + 0 (first asserted pixel in vectorized word)
8. VEC X INTENSITY MSB – 0 + 1 (second asserted pixel in vectorized word)
9. VEC X INTENSITY LSB – 0 + 1 (second asserted pixel in vectorized word)
10. ….
11. VEC X INTENSITY MSB – 0 + 55 ($56^{th}$ asserted pixel in vectorized word, worst case scenario)
12. VEC X INTENSITY LSB – 0 + 55 ($56^{th}$ asserted pixel in vectorized word)

## 5.6 DATUM Decoding Table:

The following is the full decoding table of all supported data types in EVT+0.7. It is then followed by the decoding table for each data type.

| # | DATUM NAME | Description | VALUE |
|---|---|---|---|
| 1 | TS MSB | MSB of timestamp | 00000001 |
| 2 | EVENT Y | Row address | 00000010 |
| 3 | MIXED X ON MSB | ON event, Col address + Intensity MSB | 00000011 |
| 4 | MIXED X OFF MSB | OFF event, Col address + Intensity MSB | 00000100 |
| 5 | MIXED X LSB | Intensity LSB (lower 24 bits) | 00000101 |
| 6 | EVENT X ON | ON event, Col address + TS LLSB | 00000110 |
| 7 | EVENT X OFF | OFF event, Col address + TS LLSB | 00000111 |
| 8 | VEC X ON MSB | On event, Col address + VEC MSB (8) | 00001000 |
| 9 | VEC X OFF MSB | OFF event, Col address + VEC MSB (8) | 00001001 |
| 10 | VEC X LSB | VEC LSB (24), thus 32 pixels at a time. | 00001010 |
| 11 | VEC X INTENSITY MSB | Intensity MSB for each asserted pixel in vectorized word | 00001011 |
| 12 | VEC X INTENSITY LSB | Intensity LSB for each asserted pixel in vectorized word | 00001100 |

*Table 2 Datum Decoding Table*

## 5.7 Payload Decoding Tables

| # | Data Out | Description | Bits Assigned (Notional) |
|---|----------|-------------|--------------------------|
| 1 | TS MSB CODE | 00000001 | 8 |
| 2 | TS PAYLOAD | Upper MSB of TS for current | 24 |
|   | **TS MSB DATA** | **Minimum # of Bits** | **32 Bits** |

*Table 3 Timestamp MSB*

### 5.7.1 Event Y Address

| # | Data Out | Description | Bits Assigned (Notional) |
|---|----------|-------------|--------------------------|
| 1 | EVENT Y | 00000010 | 8 |
| 2 | TS LSB DATA | Bottom 8 bits of Timestamp | 8 |
| 3 | EVENT Y PAYLOAD | Y (row) address of EBS | 16 |
|   | **EBS Data** | **Minimum # of Bits** | **32 Bits** |

*Table 4 Event Y Address*

### 5.7.2 Mixed ON Event X Address MSB

| # | Data Out | Description | Bits Assigned (Notional) |
|---|----------|-------------|--------------------------|
| 1 | MIXED X ON MSB | 00000011 | 8 |
| 2 | PIXEL X ADDRESS | X (col) address of EBS | 16 |
| 3 | Intensity Value | Intensity value MSB | 8 |
|   | **EBS Data** | **Minimum # of Bits** | **32 Bits** |

*Table 5 Mixed ON Event X Address MSB*

### 5.7.3 Mixed OFF Event X Address MSB

| # | Data Out | Description | Bits Assigned (Notional) |
|---|---|---|---|
| 1 | MIXED X OFF MSB | 00000100 | 8 |
| 2 | PIXEL X ADDRESS | X (col) address of EBS | 16 |
| 3 | Intensity Value | Intensity value MSB | 8 |
|   | **EBS Data** | **Minimum # of Bits** | **32 Bits** |

*Table 6 Mixed OFF Event X Address MSB*

### 5.7.4 Mixed Event X LSB

| # | Data Out | Description | Bits Assigned (Notional) |
|---|---|---|---|
| 1 | MIXED X LSB | 00000101 | 8 |
| 2 | Intensity Value | Intensity value MSB | 24 |
|   | **EBS Data** | **Minimum # of Bits** | **32 Bits** |

*Table 7 Mixed Event X LSB*

### 5.7.5 ON Event X Address

| # | Data Out | Description | Bits Assigned (Notional) |
|---|---|---|---|
| 1 | EVENT X ON | 00000110 | 8 |
| 2 | PIX X ADDRESS | X (col) address of EBS | 16 |
| 3 | TS LLSB | Lower LOWER 8-bits for enhanced temporal resolution (40-bits total) | 8 |
|   | **EBS Data** | **Minimum # of Bits** | **32 Bits** |

*Table 8 ON Event X Address*

### 5.7.6 OFF Event Y Address

| # | Data Out | Description | Bits Assigned (Notional) |
|---|---|---|---|
| 1 | EVENT X OFF | 00000111 | 8 |
| 2 | PIX X ADDRESS | X (col) address of EBS | 16 |
| 3 | TS LLSB | Lower LOWER 8-bits for enhanced temporal resolution (40-bits total) | 8 |
|   | **EBS Data** | **Minimum # of Bits** | **32 Bits** |

*Table 9 OFF Event Y Address*

### 5.7.7 Vectorized ON Event X Address MSB

| # | Data Out | Description | Bits Assigned (Notional) |
|---|---|---|---|
| 1 | VEC X ON MSB | 00001000 | 8 |
| 2 | PIX X ADDRESS | X (col) address of EBS | 16 |
| 3 | VEC DATA MSB | One-hot code where 1 is presence of ON event | 8 |
|   | **EBS Data** | **Minimum # of Bits** | **32 Bits** |

*Table 10 Vectorized ON Event X Address MSB*

### 5.7.8 Vectorized OFF Event X Address MSB

| # | Data Out | Description | Bits Assigned (Notional) |
|---|---|---|---|
| 1 | VEC X OFF MSB | 00001001 | 8 |
| 2 | PIX X ADDRESS | X (col) address of EBS | 16 |
| 3 | VEC DATA MSB | One-hot code where 1 is presence of OFF event | 8 |
|   | **EBS Data** | **Minimum # of Bits** | **32 Bits** |

*Table 11 Vectorized OFF Event X Address MSB*

### 5.7.9 Vectorized Event LSB

| # | Data Out | Description | Bits Assigned (Notional) |
|---|---|---|---|
| 1 | VEC X LSB | 00001010 | 8 |
| 2 | VEC DATA LSB | One-hot code where 1 is presence of an event (these can be daisy chained) | 24 |
|  | **EBS Data** | **Minimum # of Bits** | **32 Bits** |

*Table 12 Vectorized Event LSB*

### 5.7.10 Vectorized Intensity Data MSB

| # | Data Out | Description | Bits Assigned (Notional) |
|---|---|---|---|
| 1 | VEC X INTENSITY MSB | 00001011 | 8 |
| 2 | INTENSITY DATA | One-hot code where 1 is presence of ON event (these can be daisy chained) | 24 |
|  | **EBS Data** | **Minimum # of Bits** | **32 Bits** |

*Table 13 Vectorized Intensity Data MSB*

### 5.7.11 Vectorized Intensity Data LSB

| # | Data Out | Description | Bits Assigned (Notional) |
|---|---|---|---|
| 1 | VEC X INTENSITY LSB | 00001100 | 8 |
| 2 | INTENSITY DATA | Bottom LSB of 32-bit intensity data supported by standard | 8 |
| 3 | EMPTY |  | 16 |
|  | **EBS Data** | **Minimum # of Bits** | **32 Bits** |

*Table 14 Vectorized Intensity Data LSB*

# 6 Conclusion

We have described a new framework for how event-based data could be structured upon output from the camera, including attempts to 'future-proof' against potential emerging modalities of capture (mixed event mode or multiple sensing types). The authors have developed the EVT+ with the intention of providing commonality and ease of use for the growing event-based data community. The choice was made to base the standard on already available hardware and data structures in order to make EVT+ easily adoptable. Feedback, comments, and suggested changes are welcome.